\documentclass[prb,aps,twocolumn,showpacs]{revtex4}

\usepackage{graphicx}
\usepackage{amssymb}

\begin{document}

\title{Anisotropic Colossal Magnetoresistance Effects in
\boldmath{Fe$_{1-x}$Cu$_x$Cr$_2$S$_4$}}

\author{V.~Fritsch$^1$, J.~Deisenhofer$^1$, R.~Fichtl$^1$,
J.~Hemberger$^1$, H.-A.~Krug von Nidda$^1$, M.~M\"{u}cksch$^2$,
M.~Nicklas$^3$, D.~Samusi$^4$, J.~D.~Thompson$^3$, R.~Tidecks$^2$,
V.~Tsurkan$^{2,4}$, A.~Loidl$^1$}

\affiliation{$^1$Experimentalphysik V, Elektronische Korrelationen
und Magnetismus, Institut f\"{u}r Physik, Universit\"{a}t Augsburg, D-86135 Augsburg \\
$^2$Institut f\"{u}r Physik, Universit\"{a}t Augsburg, D-86135 Augsburg \\
$^3$Los Alamos National Laboratory, Los Alamos, NM 87545, USA \\
$^4$Institute of Applied Physics, Academy of Science of Moldova,
Academiei 5, MD-2028, Chisinau}

\begin{abstract}
A detailed study of the electronic transport and magnetic
properties of Fe$_{1-x}$Cu$_x$Cr$_2$S$_4$ ($x \leq 0.5$) on single
crystals is presented. The resistivity is investigated for $2 \leq
T \leq 300$~K in magnetic fields up to $140$~kOe and under
hydrostatic pressure up to 16~kbar. In addition magnetization and
ferromagnetic resonance (FMR) measurements were performed. FMR and
magnetization data reveal a pronounced magnetic anisotropy, which
develops below the Curie temperature, $T_{\mathrm{C}}$, and
increases strongly towards lower temperatures. Increasing the Cu
concentration reduces this effect. At temperatures below 35~K the
magnetoresistance, $MR = \frac{\rho(0) - \rho(H)}{\rho(0)}$,
exhibits a strong dependence on the direction of the magnetic
field, probably due to an enhanced anisotropy. Applying the field
along the hard axis leads to a change of sign and a strong
increase of the absolute value of the magnetoresistance. On the
other hand the magnetoresistance remains positive down to lower
temperatures, exhibiting a smeared out maximum with the magnetic
field applied along the easy axis. The results are discussed in
the ionic picture using a triple-exchange model for electron
hopping as well as a half-metal utilizing a band picture.
\end{abstract}


\maketitle

\section{Introduction}
Manganites, especially LaMnO$_3$ and relatives, are known for
their unusual transport and magnetic properties since more than 50
years.\cite{jon50,san50} However, the appreciation and intensive
interest is a recent development, which started with the giant
magnetoresistance (actually named colossal magnetoresistance, CMR)
in thin films of La$_{2/3}$Ba$_{1/3}$MnO$_3$, published by von
Helmolt {\it et al.} in 1993,\cite{hel93} even though a negative
magnetoresistance of nearly 20 \% was discovered in bulk
La$_{0.69}$Pb$_{0.31}$MnO$_3$ by Searle and Wang already in
1970.\cite{sea70} Soon after the onset of the renewed interest in
these materials, it was realized that the theoretical framework
used in the past to understand the manganites' behaviour does not
survive a quantitative analysis.\cite{mil95,sal01} The complexity
of the problem led to the perception that manganites are
prototypical for correlated electron systems, where spin, charge
and orbital degrees of freedom are strongly coupled. These
couplings lead to a failure of the classical approach, which
neglects some interactions for simplification, and opens the way
for a complete range of new physics. As a consequence the
experimental and theoretical studies of manganites and related
compounds give the unique opportunity of getting a deeper
understanding of the fundamental physics responsible for phenomena
like colossal magnetoresistance or high-temperature
superconductivity.

Looking for new materials exhibiting a CMR effect, the
substitution of oxygen with the isoelectronic sulphur seems to be
a promising way.\cite{ram97b} Magnetoresistance effects in some
chalcogenide spinels were reported previously by Watanabe
\cite{wat73} and Ando \cite{and79}. An elaborately review about
this is given in Ref.~\onlinecite{sta82}. Since the CMR is
associated with a double-exchange mechanism, the rediscovery of a
CMR effect in the chalcospinel FeCr$_2$S$_4$,\cite{ram97a} which
is neither oxide nor perovskite, opened a wide field for the
further exploration and exploitation of magnetoresistance effects.

FeCr$_2$S$_4$ is a ferrimagnetic semiconductor, crystallizing in
the normal spinel structure, where the Cr ions occupy the
octahedral and the Fe ions the tetrahedral sites.\cite{haa67} The
Fe- and the Cr-sublattices are aligned antiparallel in the
ferrimagnetic state. In single-crystalline FeCr$_2$S$_4$ the Curie
temperature is $T_C = 167$~K and around $T_{\mathrm{C}}$ a
negative magnetoresistance is observed.\cite{ram97a} Doping with
non-magnetic Cu on the Fe site, Fe$_{1-x}$Cu$_x$Cr$_2$S$_4$ ($x
\leq 0.5$), shifts the Curie temperature upwards accompanied by a
decreasing magnetoresistance without changing substantially the
magnetic properties.\cite{haa67}

Polycrystalline samples of Fe$_{1-x}$Cu$_x$Cr$_2$S$_4$ were first
synthesized in the fifties and sixties of the last century.
\cite{hah56} To explain the physical properties two competing
models with different valences of the involved ions were proposed.
Lotgering {\it et al.}\cite{lot69} developed a model considering a
monovalent Cu$^+$-ion over the whole concentration range, while
Goodenough \cite{goo69} postulated divalent Cu$^{2+}$ for the
concentration range $0.5 < x \leq 1$. Furthermore the existence of
monovalent S$^-$ was discussed at these times.\cite{lot69}

M\"{o}{\ss}bauer-spectroscopy studies reveal divalent Fe$^{2+}$ ions in
FeCr$_2$S$_4$, but trivalent Fe$^{3+}$ in
Fe$_{0.5}$Cu$_{0.5}$Cr$_2$S$_4$.\cite{haa68a,che99} X-ray
photoelectron-spectroscopy measurements show that Cu is monovalent
in Fe$_{0.5}$Cu$_{0.5}$Cr$_2$S$_4$ and in CuCr$_2$Se$_4$, which
means it is in a $3d^{\mathrm{10}}$ state \cite{tsu00}.
NMR-measurements and band-structure calculations lead to the same
conclusion for the Cu valence in CuIr$_2$S$_4$.\cite{mat97,rad02}
All samples under investigation in this study were prepared as
described in Ref.~\onlinecite{tsu00} and found to contain only
divalent S. Therefore Cu existing only in the non-magnetic
$3d^{\mathrm{10}}$ state and divalent S only is assumed. The later
discussion adopts this assumption.

\section{Experimental Methods}
Single crystals of Cu-substituted FeCr$_2$S$_4$ were grown by the
chemical transport-reaction method from polycrystalline material
obtained by a solid-state reaction. In this paper samples with
Cu-concentrations $x = 0.05$, $0.1$, $0.2$, $0.3$, $0.4$, and
$0.5$ are studied.

The X-ray diffraction measurements were performed with a Stoe
X-ray diffractometer. Single crystals were powdered and
diffraction spectra were taken from $35^{\circ}$ to $130^{\circ}$
and analyzed with the Visual X$^{\mathrm{POW}}$ software.

The magnetic properties were measured using a superconducting
quantum interference device (SQUID) magnetometer (Quantum Design)
in the temperature range $1.8 \leq T \leq 400$~K in external
fields up to 70~kOe. In addition ferromagnetic (or better
ferrimagnetic) resonance (FMR) measurements were carried out at
X-band frequencies (9.4~GHz) with a Bruker ELEXSYS E500-CW
spectrometer using a continuous Helium gas-flow cryostat (Oxford
Instruments) for temperatures $4.2\leq T \leq 300\mathrm{~K}$. For
the FMR experiments thin polished disks prepared in $(110)$ plane
orientation with about 1~mm diameter and 0.05~mm thickness were
used.

The electrical resistivity was measured in an Oxford
$^{\mathrm{4}}$He cryostat equipped with a superconducting magnet
capable of magnetic fields up to 16~T. Conventional dc four-point
techniques were used with currents between 0.5 and 500 $\mu$A at
temperatures $2 \leq T \leq 300\mathrm{~K}$. Gold wire with a
diameter of $25~\mu$m and silver paint were used to prepare the
electrical contacts. The contact resistance was always between
$20$ and $70~\Omega$. To prevent problems occurring due to aging
of the contacts, leading to a contact resistance several orders of
magnitudes higher, the measurements were performed immediately
after preparing the contacts.

Hydrostatic pressure was produced in a conventional Be-Cu clamp
type cell using fluorinert\texttrademark\/ as a pressure medium.
The pressure at low temperatures was determined from the shift of
the inductively measured $T_{\mathrm{C}}$ of a small piece of
lead, located in immediate proximity to the sample. The width of
the superconducting transition of Pb did not exceed 15~mK,
indicating good hydrostatic conditions and providing an estimate
of the pressure-measurement uncertainty, $\pm0.4$~kbar. The
pressure at room temperature was determined from the pressure
dependence of the resistivity of a manganin wire placed inside the
cell.

\section{Experimental Results}
\subsection{X-ray Diffraction}

\begin{figure}
\includegraphics[clip,width=80mm]{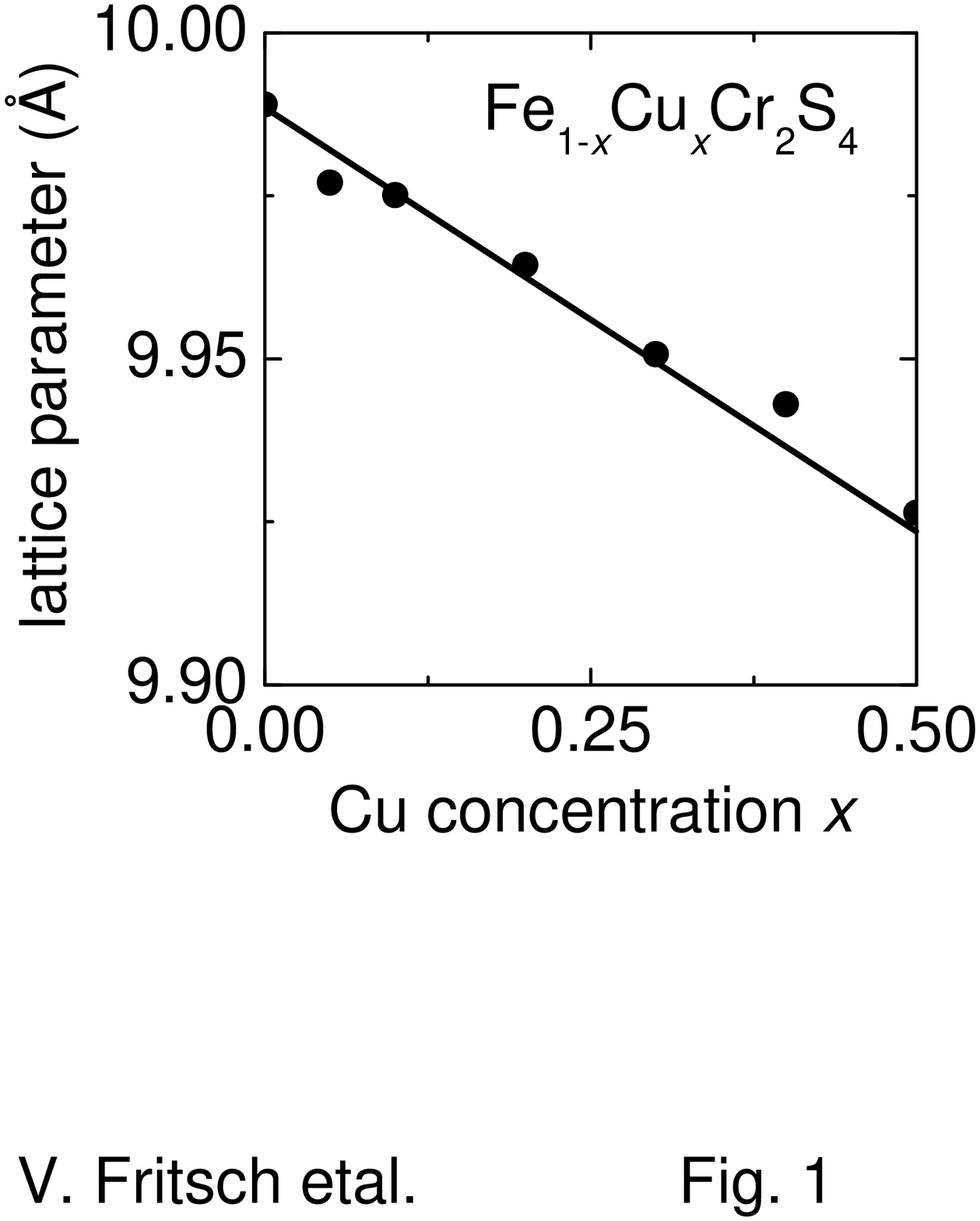}
\caption{Cubic lattice parameter vs Cu concentration $x$ in
Fe$_{1-x}$Cu$_x$Cr$_2$S$_4$. \label{gitterkonstante}}
\end{figure}

In FeCr$_2$S$_4$ the substitution of Fe by Cu leads to a linear
dependence of the lattice parameter of the cubic spinel structure
on the Cu concentration following Vegard's law, as shown in
figure~\ref{gitterkonstante}. In addition, the X-ray studies of
powdered single crystals confirmed single-phase material with no
detectable parasitic phases.

\subsection{Magnetization}
\begin{figure}
\includegraphics[clip,width=80mm]{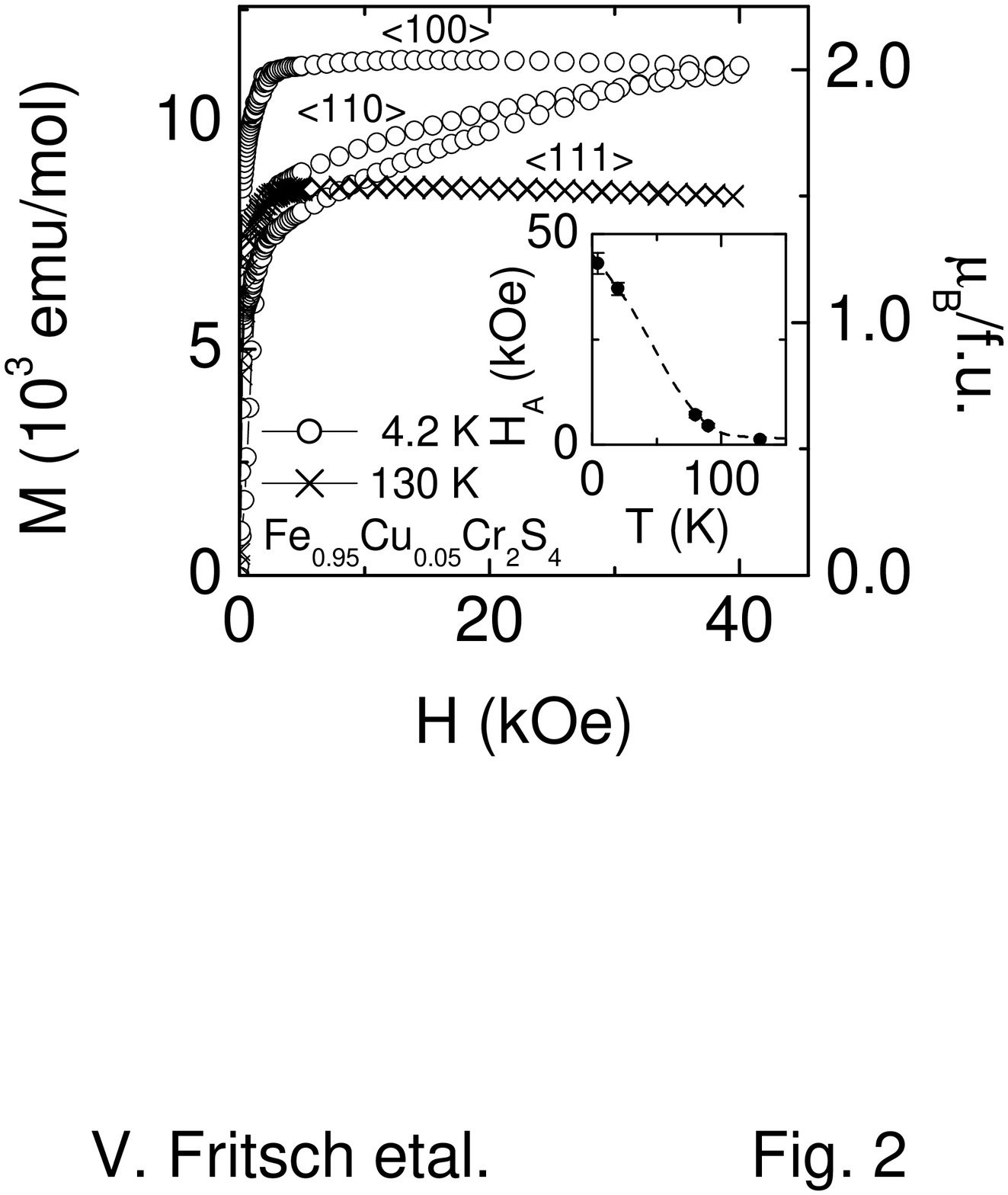}
\caption{Magnetization of Fe$_{0.95}$Cu$_{0.05}$Cr$_2$S$_4$ vs.
magnetic field at $T = 4.2$~K ($\bigcirc$) and $T = 130$~K
($\times$, here all crystallographic orientations are nearly
coincide) respectively. Inset: Anisotropy field $H_A$ vs.
temperature. The dashed line is to guide the eyes. \label{M(H)}}
\end{figure}

\begin{table}
\caption{Curie temperature $T_{\mathrm{C}}$ of
Fe$_{1-x}$Cu$_x$Cr$_2$S$_4$, determined by magnetization
measurements, and electrical resistivity, $\rho$, at room
temperature ($T = 290$~K) for specimens of different Cu
concentrations, $x$. \label{TC}}
\begin{tabular}{ccc}
 \hspace{2cm} & \hspace{2cm} &\hspace{4cm} \\
 \hline \hline
 $x$   &   $T_{\mathrm{C}}$ (K $\pm$ 0.5)    & $\rho$(290~K) (m$\Omega$cm $\pm 10\%$) \\
 \hline
 0      &  167     &   236 (Ref.~\onlinecite{tsu01b}) \\
 0.05   &  182     &   79.2    \\
 0.1    &  197     &   8.2    \\
 0.2    &  215     &   10.1    \\
 0.3    &  232     &   11.6    \\
 0.4    &  236     &   14.9   \\
 0.5    &  275     &   26.8    \\
 \hline \hline
\end{tabular}
\end{table}

The Curie temperatures $T_{\mathrm{C}}$ of
Fe$_{1-x}$Cu$_x$Cr$_2$S$_4$ are listed in table~\ref{TC}, as
determined by the kink-point-method \cite{arr71} from
magnetization measurements, and the room temperature resistivity,
which will be discussed in section~\ref{rhoroom}. The Curie
temperature $T_{\mathrm{C}}$ increases with the Cu-concentration
$x$. The same trend has been observed for polycrystalline
samples,\cite{haa67} though for higher Cu concentrations
$T_{\mathrm{C}}$ remains at a lower value in single crystals.
Figure~\ref{M(H)} shows the magnetization, $M$, for
Fe$_{0.95}$Cu$_{0.05}$Cr$_2$S$_4$ versus the magnetic field,~$H$,
at $T = 4.2$~K and $T = 130$~K, respectively. At $T = 4.2$~K the
magnetic anisotropy is clearly observed. For the easy
magnetization axis $\langle 100\rangle $ the saturation  is
already reached at $2$~kOe whereas for the hard axis $\langle
111\rangle $ and the intermediate axis $\langle 110\rangle $
saturation only occurs at $43$~kOe. The temperature dependence of
the anisotropy field $H_A$, defined by the magnetic field where
saturation is reached for all three directions, is shown in the
inset of figure~\ref{M(H)}. It decreases monotonically with
increasing temperature and vanishes at $T_{\mathrm{C}}$.

\subsection{FMR-Measurements}
\begin{figure}
\includegraphics[clip,width=80mm]{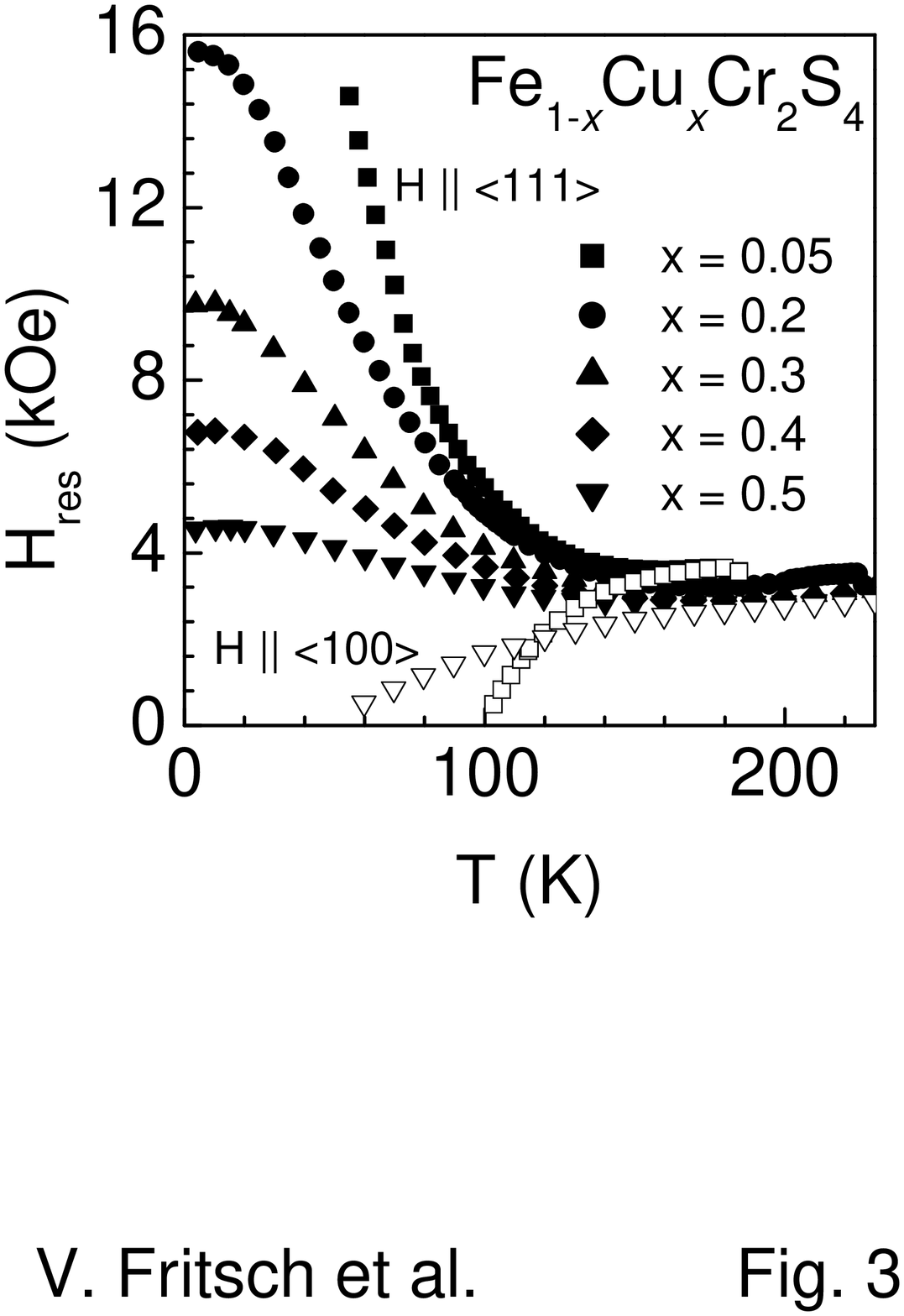}
\caption{Resonance field $H_{res}$ of Fe$_{1-x}$Cu$_x$Cr$_2$S$_4$
of the FMR spectrum as a function of the temperature for the
magnetic field applied parallel to the hard axis $\langle
111\rangle $ (closed symbols). Additionally the resonance field
$H_{res}$ for $x = 0.05$ and $0.5$ with the magnetic field applied
in the easy direction $\langle 100\rangle $ are shown (open
symbols). The anisotropy of $H_{res}$ reflects the
magnetocrystalline anisotropy in the system. \label{esr}}
\end{figure}

For a more detailed analysis of the magnetic anisotropy we
performed ferromagnetic resonance (FMR) measurements, which will
be published in a separate paper. Here we confine ourselves to the
presentation of one illustrative result, which nicely reflects the
evolution of the anisotropy with increasing Cu concentration and
can be explained on the base of the FMR results published recently
for FeCr$_2$S$_4$ single crystals.\cite{tsu01a} For the samples
under investigation ($0 < x \leq 0.5$) the FMR line exhibits an
analogous behavior to the pure compound $x=0$. Figure~\ref{esr}
shows the temperature dependence of the resonance field $H_{\rm
res}$ for several Cu concentrations below the Curie temperature.
The static magnetic field was applied along an $\langle 111\rangle
$ or $\langle 100\rangle $ axis within the $(110)$ plane of the
disk-shaped samples and the magnetic microwave field was applied
perpendicular to the plane. This geo\-me\-try allows measurements
at different orientations of the static field in the plane without
change of the demagnetization contributions to the resonance
condition.\cite{kit47,kit48} Just below the Curie temperature the
resonance field $H_{\rm res}$ is approximately isotropic given by
the Larmor frequency $\nu = \gamma H_{\rm res}$, with the
microwave frequency $\nu$ and the gyromagnetic ratio $\gamma$
determined by the g-values of the two sublattices.\cite{gur96}
With decreasing temperature one observes first a slight shift to
smaller fields due to the demagnetization but then a strongly
anisotropic behavior appears. For the magnetic field applied along
the easy $\langle 100\rangle $ axis, the resonance line shifts to
low fields and disappears at a finite temperature as shown
exemplarily for $x=0.05$ and $x=0.5$. For the field applied
parallel to the hard $\langle 111\rangle $ axis, the resonance
field shifts to higher fields. A similar shift to higher fields is
observed for orientation along the intermediate $\langle
110\rangle $ axis (not shown in figure~\ref{esr}). The maximum
shift at low temperatures decreases with increasing Cu
concentration.

This result is directly related to the decrease of the magnetic
anisotropy. Neglecting the demagnetization effects, which turn out
to be small compared to the anisotropy field at low temperatures
\cite{tsu01a} and taking into account only the first-order cubic
anisotropy field $H_{\rm A}=K_1/M$, where $K_1$ is the first-order
cubic anisotropy constant, the resonance conditions read for
$\langle 100\rangle $ and $\langle 111\rangle $ orientation,
respectively:\cite{gur96}

\begin{eqnarray}\label{reso}
\frac{\nu}{\gamma} = H_{\rm res}^{100} + 2H_{\rm A}, \nonumber \\
\frac{\nu}{\gamma} = H_{\rm res}^{111} - \frac{4}{3}H_{\rm A}.
\end{eqnarray}

Hence, the resonance shift $H_{\rm res}-\nu/\gamma$ from the
Larmor frequency is proportional to the anisotropy field. For
$H||\langle 100\rangle $ the shift is negative and the resonance
disappears at zero field. However, the shift is positive for
$H||\langle 111\rangle $ and can be followed down to lowest
temperatures, only limited by the field range, which is accessible
to the electromagnet. For this reason we can directly compare the
temperature dependence of the anisotropy field calculated from the
magnetization measurements for $x=0.05$ (inset of
figure~\ref{M(H)}) with the temperature dependence of the FMR
shift and use the results from FMR to determine the anisotropy
field for all Cu concentrations. This clearly indicates the
continuous decrease of the magnetic anisotropy with increasing Cu
concentration.

\subsection{Electrical Resistivity}
\begin{figure}
\includegraphics[clip,width=80mm]{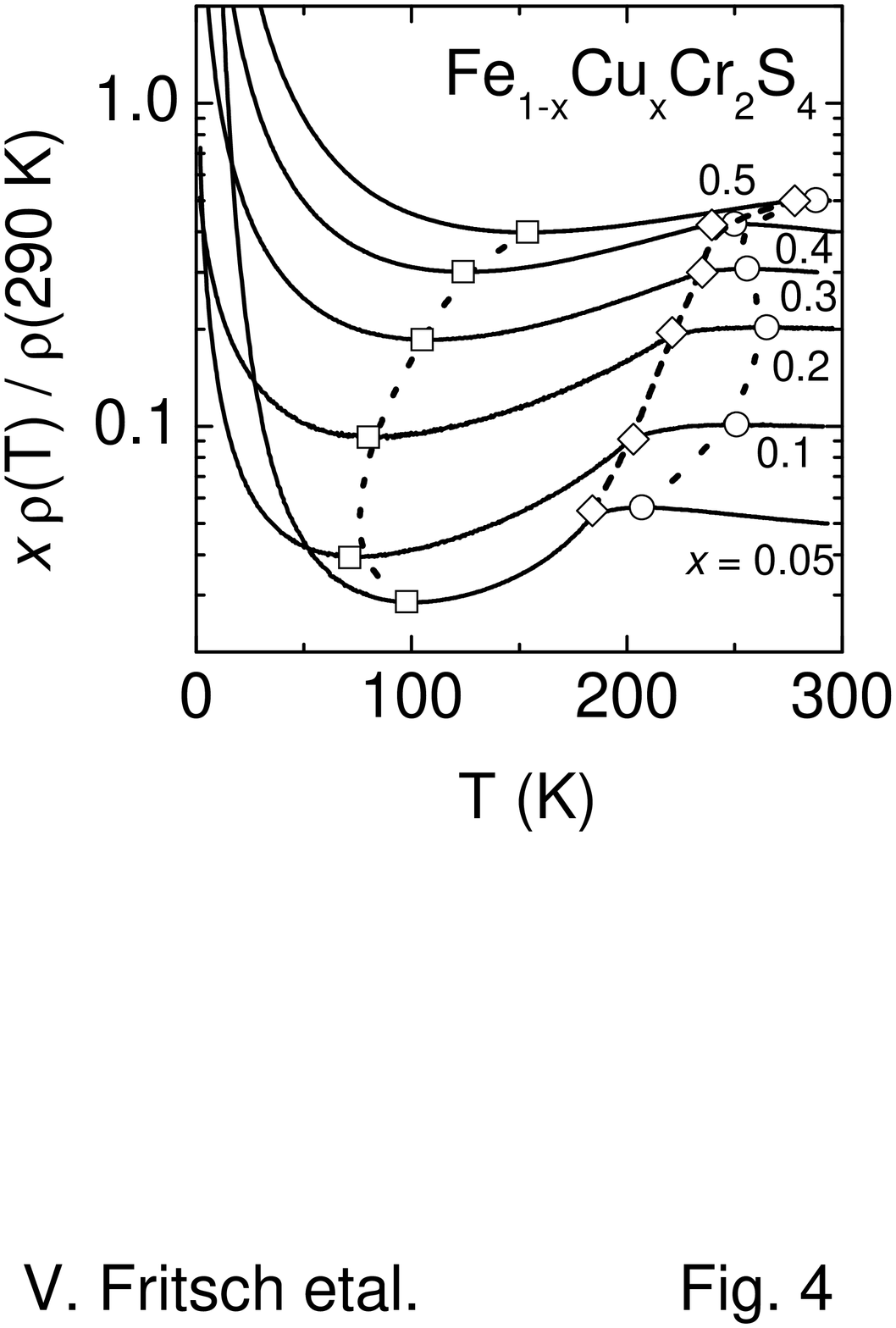}
\caption{Electrical Resistivity of Fe$_{1-x}$Cu$_x$Cr$_2$S$_4$
normalized  at $T = 290$~K and  multiplied with the
Cu-concentration $x$. The Cu concentrations $x$ are indicated in
the figure. Additionally the Curie temperatures $T_{\mathrm{C}}$
({\large $\diamond$}, dashed line) and the positions of the local
minima ($\square$, dotted line) and maxima ({\large $\circ$},
dotted line) are given. The current was applied along the $\langle
110\rangle $-direction.\label{Rnorm}}
\end{figure}

Figure~\ref{Rnorm} shows a semi-logarithmic plot of the
resistivity of Fe$_{1-x}$Cu$_x$Cr$_2$S$_4$ normalized by the
room-temperature resistivity and multiplied with the Cu
concentrations $x = 0.05$, $0.1$, $0.2$, $0.3$, $0.4$, and $0.5$,
to enable the identification of the different concentrations. The
absolute values of the resistivity at room temperature are
summarized in table~\ref{TC}. One should keep in mind that such
absolute values are given with a large uncertainty. The given
error of $10\%$ is the error due to the determination of the
geometric factor. A similar order of magnitude and tendency of
concentration dependence of the values given here was observed in
single crystals by Haacke and Beegle.\cite{haa68b}

The resistivity of Fe$_{1-x}$Cu$_x$Cr$_2$S$_4$ ($x \leq 0.5$)
exhibits a non-monotonic behavior with a local maximum slightly
above and a broad minimum below $T_{\mathrm{C}}$. The resistivity
increases strongly at low temperatures, indicating the insulating
ground state of the system.  The existence of the local extrema is
in agreement with the results in
FeCr$_2$S$_4$.\cite{ram97a,wat73,tsu01b}

\begin{figure}
\includegraphics[clip,width=80mm]{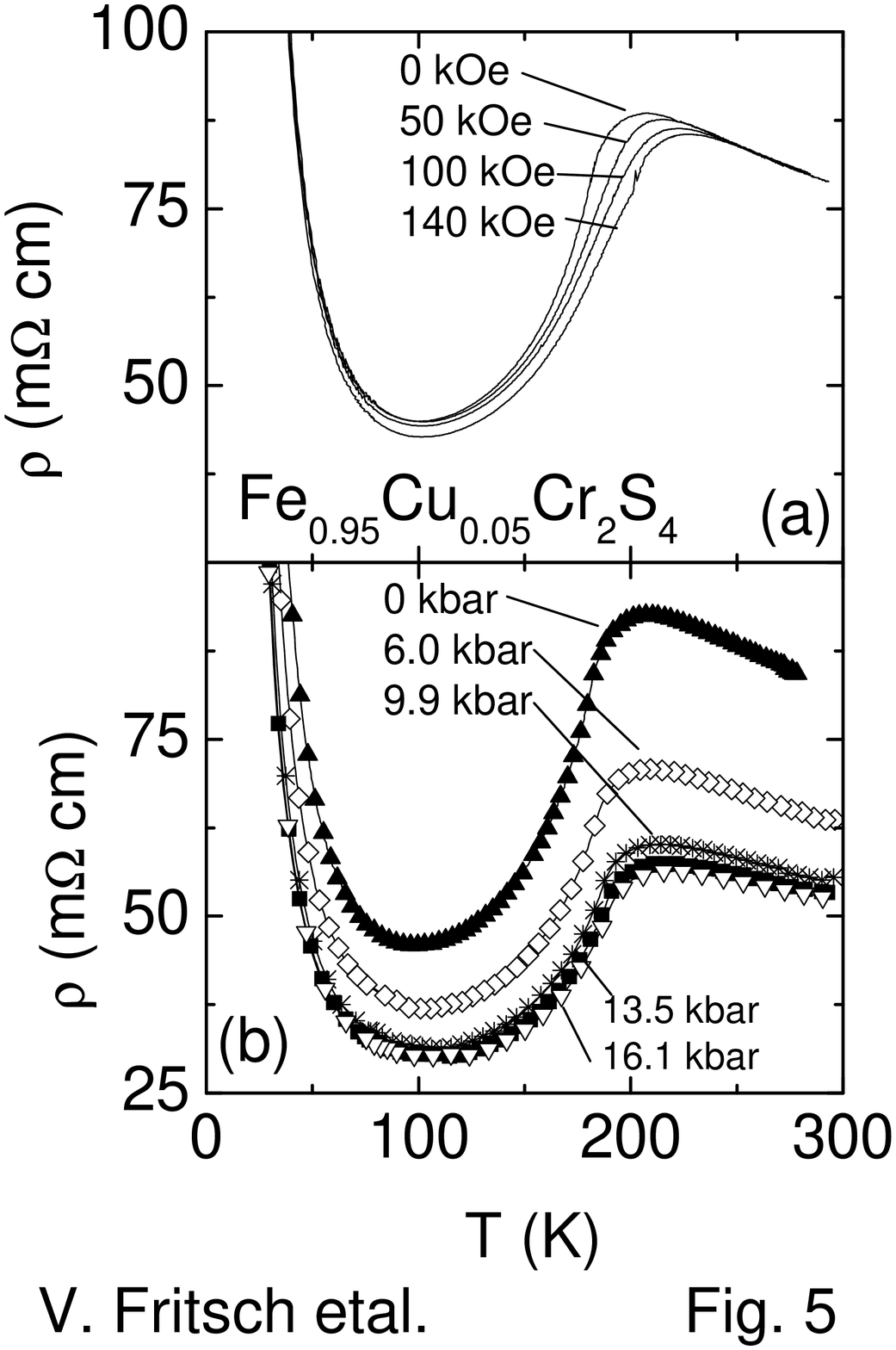}
\caption{(a) Resistivity of Fe$_{0.95}$Cu$_{0.05}$Cr$_2$S$_4$ near
$T_{\mathrm{C}}$ in magnetic fields up to 140~kOe. The magnetic
field is applied in $\langle 111\rangle $-direction with the
current in $\langle 110\rangle $-direction. (b) Resistivity of
Fe$_{0.95}$Cu$_{0.05}$Cr$_2$S$_4$ under hydrostatic pressure.
\label{Fe50V}}
\end{figure}

The resistivity of Fe$_{0.95}$Cu$_{0.05}$Cr$_2$S$_4$ is plotted in
figure~\ref{Fe50V}~(a) for different magnetic fields, 0, 50, 100
and 140~kOe. The magnetic field is applied along the hard axis
($\langle 111\rangle $-direction), the current is applied in
$\langle 110\rangle$-direction. The maximum in the vicinity of the
Curie temperature $T_{\mathrm{C}}$ slightly shifts to higher
temperatures, while the minimum remains at a constant temperature
with increasing magnetic field. The concentrations $x = 0.1$, 0.2,
0.3, and 0.5 show a similar dependence on magnetic field.

The resistivity of Fe$_{0.95}$Cu$_{0.05}$Cr$_2$S$_4$ was also
measured under hydrostatic pressure. In figure~\ref{Fe50V}~(b) the
resistivity of Fe$_{0.95}$Cu$_{0.05}$Cr$_2$S$_4$ for different
pressures up to $16.1$~kbar is shown. Under a pressure of
$16.1$~kbar the resistivity is reduced by $37~\%$ at room
temperature. The minimum as well as the local maximum are shifted
to higher temperatures (see figure~\ref{vvsp}~(b) for
$T_{\mathrm{max}}$ and $\frac{dT_{\mathrm{max}}}{dp}$).

\section{Discussion}
\subsection{The ionic picture: Triple exchange model}\label{model}
The system Fe$_{1-x}$Cu$_x$Cr$_2$S$_4$ can be divided in two
different concentration regimes $0 \leq x \leq 0.5$ and $0.5 < x
\leq 1$ with different physical properties. The concentration
range $0.5 < x \leq 1$ will be treated in a forthcoming paper.

\begin{figure}
\includegraphics[clip,width=80mm]{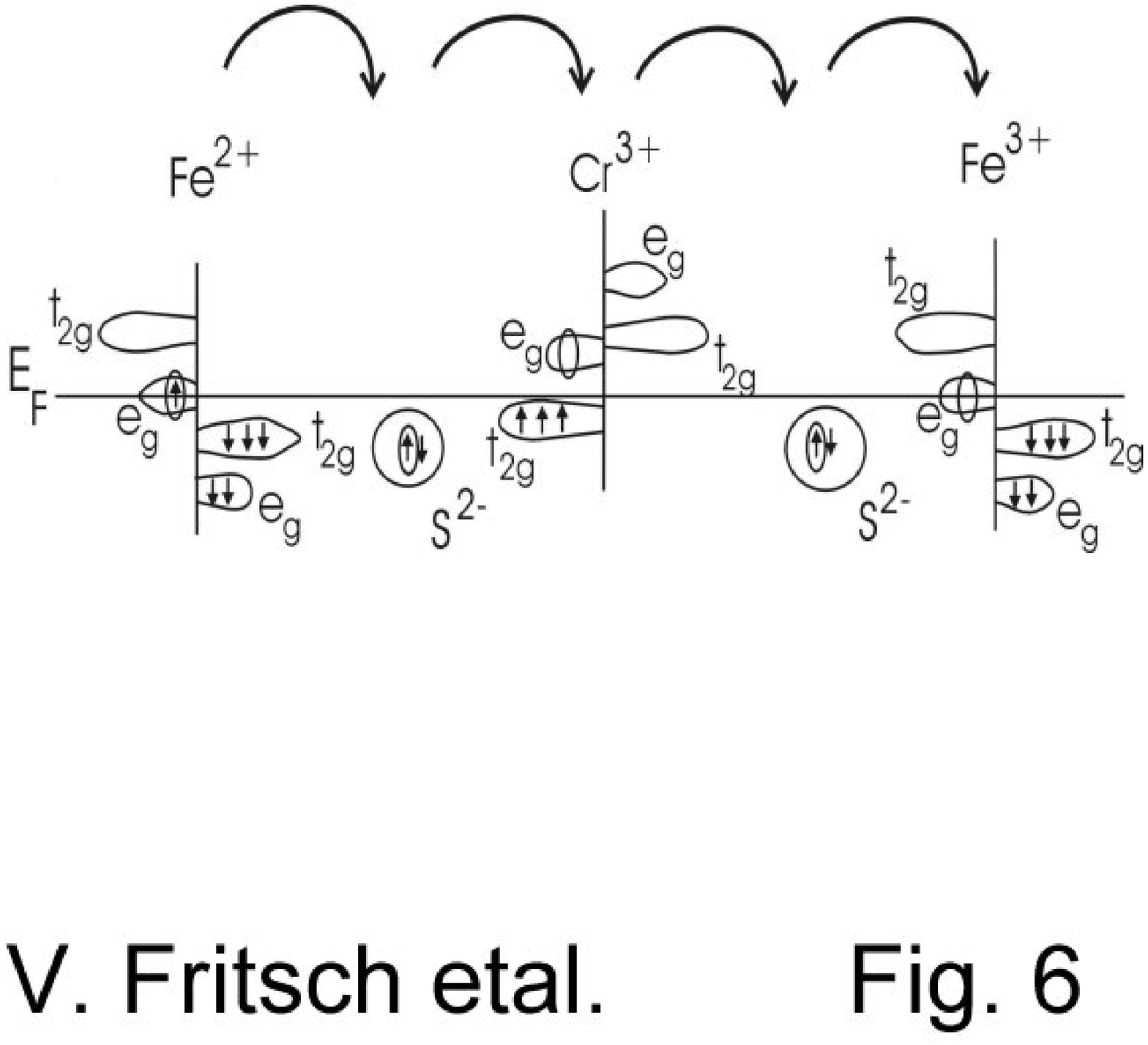}
\caption{Illustration of the triple-exchange between Fe$^{2+}$ and
Fe$^{3+}$ via S and Cr. The rough position of the bands is adopted
from the band structure calculations of Park {\it et
al.}\cite{par99} The mobile electrons and the empty states, into
which they are hopping, are circled. For details see
text.\label{austausch}}
\end{figure}

In the region $x \leq 0.5$ the valences of the ions can be
described by the formula
\begin{equation}
\mathrm{Fe}_{1-2x}^{2+}\mathrm{Fe}_{x}^{3+}\mathrm{Cu}_{x}^{+}\mathrm{Cr}_{2}^{3+}\mathrm{S}_{4}^{2-}.
\end{equation}
This description was already given by Lotgering {\it et
al.}\cite{lot69} and Goodenough.\cite{goo69} As a conduction
mechanism Palmer and Greaves proposed a triple-exchange
model.\cite{pal99} In this model the electrical conduction is
established via hopping between Fe$^{2+}$ and Fe$^{3+}$. An
illustration is given in figure~\ref{austausch}. Fe$^{2+}$ has six
$3d$-electrons, where the sixth electron is located in the
$e_g$-band with the spin antiparallel to the spins of the other
five electrons of Fe and parallel to the Cr-moments, which define
the direction of the magnetization. The single electron in the
Fe's spin-up $e_g$-band hops with an exchange mechanism, similar
to the well-known double-exchange,\cite{zen53} via a $p$-orbital
of the sulphur to Cr, providing an additional electron on the Cr
site leading to an intermediate Cr$^{2+}$ state. From there it
proceeds via the second S to the Fe$^{3+}$, changing the valence
to Fe$^{2+}$. Because of its antiparallel alignment to the
remaining $d$-electron spins of the Fe, the spin of the hopping
electron is parallel to the spin of the electrons in the Cr
$3d$-band.\cite{pal99}

The observed temperature and magnetic field dependence of the
resistivity for $0 < x < 0.5$ can be explained by the
triple-exchange model. In the paramagnetic region above
$T_{\mathrm{C}}$ semi-conducting behaviour due to thermal
activated hopping is observed. At $T_{\mathrm{C}}$ the system
enters the magnetically ordered state and the Cr and Fe spins are
aligned antiparallel, stimulating the hopping via the
triple-exchange mechanism and leading to a positive temperature
slope of the resistivity. For the absolute values of the
resistivity one would expect a minimum at a Cu concentration $x
\approx \frac{1}{3}$, where an equal amount of Fe$^{2+}$ and
Fe$^{3+}$ exists. The values for the resistivity given in
table~\ref{TC} show a broad minimum between $x = 0.1$ and $0.3$.
This is an indication, that the system cannot be described by an
pure ionic picture only. Thus, in the next section a description
in a band picture will be given.

The attempts to fit the low-temperature increase of the
resistivity with an Arrhenius- ($\rho \propto
\exp\left(\frac{T_0}{T}\right)$) or a variable-range hopping law
($\rho \propto \exp\left(\left(
\frac{T_0}{T}\right)^{\frac{1}{4}}\right)$) failed. The rise of
the resistivity is weaker than either a simple Arrhenius- or
variable-range hopping law and probably cannot be explained by
only a single mechanism alone. In the whole temperature regime,
variable range hopping is assumed to be the relevant conduction
mechanism. But below the ordering temperature the triple exchange
enhances the conductivity compared to the simple variable-range
hopping process, correlated with the magnetic anisotropy. Also in
the ordered phase, there might be additional contributions to the
resistivity from magnon scattering which increases with increasing
temperature.

\subsection{The band picture: Fe$_{1-x}$Cu$_x$Cr$_2$S$_4$ as a
half metal} We assume the Fermi-edge to be located within the Fe
spin-up $e_g$-band, as it is shown in figure~\ref{austausch}. This
assumption is supported by band-calculations of Park {\it et al.},
who describe Fe$_{1-x}$Cu$_x$Cr$_2$S$_4$ as a half-metallic
ferromagnet.\cite{par99} The half-metallic ferromagnetic state is
realized, if all spins are fully polarized forming one metallic
and one insulating band.\cite{gro83} From their calculations Park
{\it et al.} expected a metallic ground state for FeCr$_2$S$_4$.
The metallic ground state is changed by Coulomb interactions
splitting the Fe $e_g$ band, leading to a
Mott-insulator.\cite{par99} In addition this splitting is
supported by the Jahn-Teller effect \cite{fei82}, which is
peculiar to Fe$^{2+}$ ions and shown by M\"{o}ssbauer
experiments.\cite{spe72}. At higher temperatures near
$T_{\mathrm{C}}$ the thermal activation is high enough to overcome
the band splitting, which leads to the observed positive
temperature gradient in the resistivity below $T_{\mathrm{C}}$.
Above $T_{\mathrm{C}}$ the spins are not ordered anymore and a
simple hopping conductivity is established.

Substituting Fe by Cu empties the Fe$^{2+}$ $e_g$ spin-up band
and, thus, destroys the band splitting, which explains the strong
decrease of the resistivity in the concentration range up to
$10~\%$.\label{rhoroom} Further substitution of Fe with Cu empties
the Fe$^{2+}$ $e_g$ spin-up band, reducing the number of charge
carriers and, thus, leads to an eventual increase of the
resistivity with increasing $x$.

At $x = 0.5$ all Fe ions should be trivalent and an insulating
ground state is found (although Park {\it et al.} assume
Cu$^{2+}$).\cite{par99} Nevertheless, in the region below
$T_{\mathrm{C}}$ a positive temperature gradient of the
resistivity is found. To understand this, one has to look on the
concentration range $x \geq 0.5$. Here we assume a double-exchange
mechanism between Cr$^{3+}$ and Cr$^{4+}$ via S, as proposed by
Lotgering {\it et al.}\cite{lot64} Slight off-stoichiometries in
Fe$_{0.5}$Cu$_{0.5}$Cr$_2$S$_4$ can lead to either that not all
Fe$^{2+}$-ions are changed completely to Fe$^{3+}$ or that already
at concentrations $x < 0.5$ Cr$^{3+}$-ions start to be turned in
Cr$^{4+}$ and, thus, give the possibility to process
double-exchange in the ordered regime below $T_{\mathrm{C}}$.

Cu$^{+}$ is in $3d^{10}$-state and therefore has a closed
$d$-shell. That is why in the ionic picture Cu is not supposed to
contribute to the conductivity. In the band picture the $t_{2g}$-
and $e_g$-band are completely filled, thus also in this case no
contribution to the conductivity is expected.

\subsection{Influence of the magnetic field}
\begin{figure}
\includegraphics[clip,width=80mm]{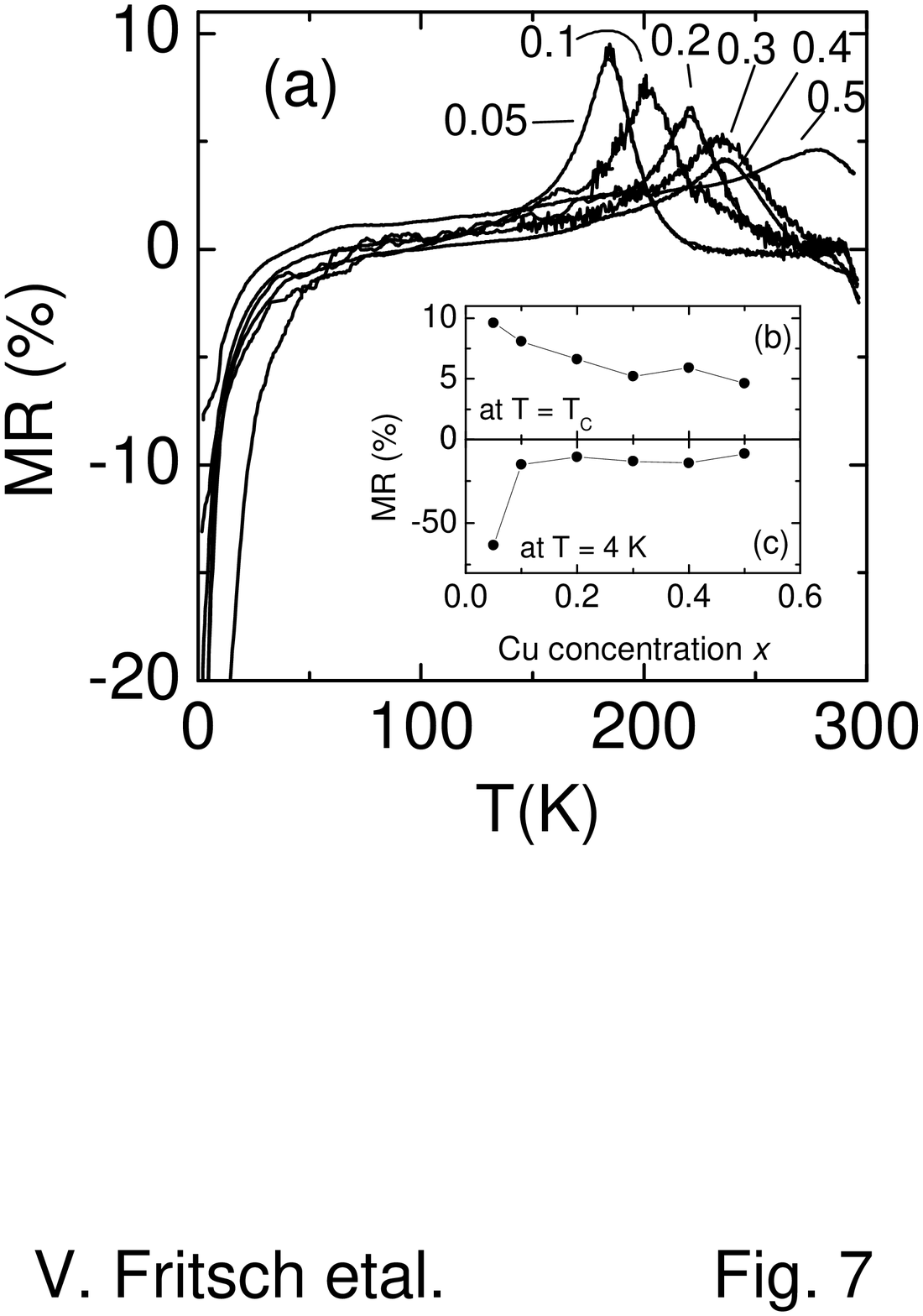}
\caption{(a): Magnetoresistance $MR = \frac{\rho (0 \mathrm{kOe})
- \rho (50 \mathrm{kOe})}{\rho (0 \mathrm{kOe})}$ of
Fe$_{1-x}$Cu$_x$Cr$_2$S$_4$ for the concentrations $x = 0.05$,
$0.1$, $0.2$, $0.3$ and $0.5$. The magnetic field is applied in
$\langle 111\rangle $-direction, the current is applied in the
$\langle 110\rangle $-direction. (b): maximum of the
magnetoresistance at the Curie temperature $T_{\mathrm{C}}$ in a
magnetic field of $H = 50$~kOe; (c): value of the
magnetoresistance at $4$~K in a magnetic fields $H = 50$~kOe vs.
Cu-concentration $x$. \label{MR}}
\end{figure}

The magnetic order is anisotropic due to a strong spin-orbit
coupling of the tetrahedral Fe$^{2+}$ ions in the $3d^{\mathrm
6}$-state.\cite{fei82,hoe72,gol76} The $6^{th}$ $d$-electron
located in the $e_g$-band (see figure~\ref{austausch}) perturbs
the symmetry of the charge distribution. This leads to a preferred
orientation of the orbitals and with the spin-orbit-coupling to
the observed magnetic anisotropy.

\begin{figure}
\includegraphics[clip,width=80mm]{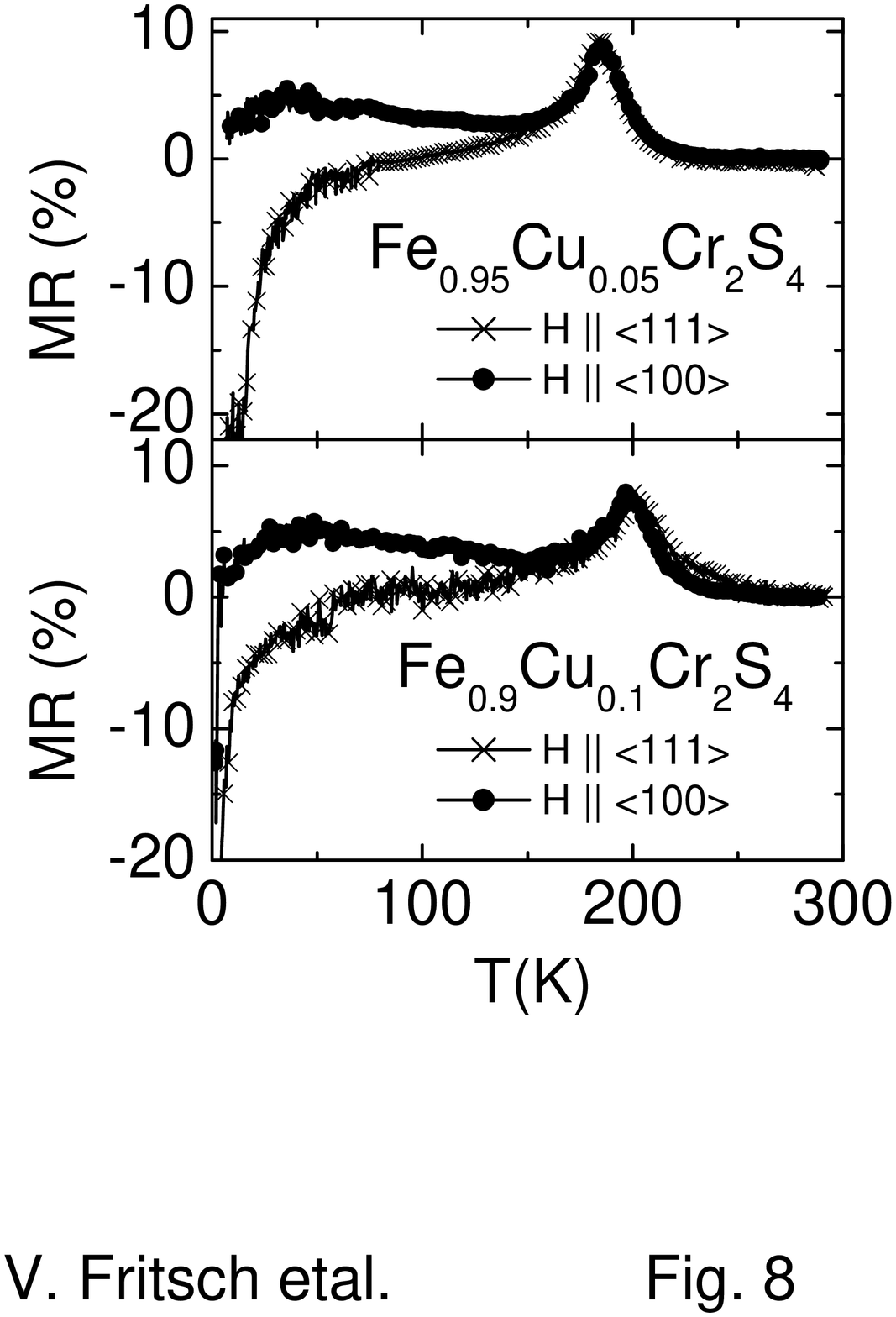}
\caption{Magnetoresistance $MR = \frac{\rho (0 \mathrm{kOe}) -
\rho (50 \mathrm{kOe})}{\rho (0 \mathrm{kOe})}$ of
Fe$_{0.95}$Cu$_{0.05}$Cr$_2$S$_4$ (upper frame) and
Fe$_{0.9}$Cu$_{0.1}$Cr$_2$S$_4$ with the magnetic field applied
along the easy axis $\langle 100\rangle $ ($\bullet$) and the hard
axis $\langle 111\rangle $ ($\times$). The current was applied in
the $\langle 110\rangle $-direction always.\label{MReasy}}
\end{figure}

In figure~\ref{MR}~(a) the magnetoresistance $MR := \frac{\rho (0
\mathrm{kOe}) - \rho (50 \mathrm{kOe})}{\rho (0 \mathrm{kOe})}$ of
Fe$_{1-x}$Cu$_x$Cr$_2$S$_4$ is displayed. Note, that in our
definition $MR > 0$ if $\rho (H) < \rho (0)$. For all
concentrations the field was applied along the hard axis $\langle
111\rangle $. As the magnetic field aligns the spins, the
triple-exchange is enhanced and the conductivity grows. This
enhancement is most pronounced at $T_{\mathrm{C}}$ due to the
onset of spontaneous order and decreases to lower temperatures. At
the Curie temperature $T_{\mathrm{C}}$ a peak arises, which was
theoretically predicted in metals \cite{cam82} and is smeared out
with increasing Cu-concentration. The maximum of the
magnetoresistance vs the Cu-concentration $x$ is drawn in
figure~\ref{MR}~(b). It drops from $9.6 \%$ at $x = 0.05$ to $4.6
\%$ at $x = 0.5$. In the region between 100 and 35~K the
magnetoresistance changes its sign and its absolute value grows up
to $63 \%$ at $T = 4$~K for $x = 0.05$. The values of the
magnetoresistance at $T = 4$~K in dependence of the
Cu-concentration $x$ are plotted in figure~\ref{MR}~(c). With
increasing Cu concentration the magnitude of the magnetoresistance
is reduced from $63 \%$ at $x = 0.05$ to $8.4 \%$ at $x = 0.5$.
Using the idea of triple exchange, the last results indicate that
obviously the magnetic field, applied along the hard axis, leads
to a weak distortion of the $e_g$-orbital of Fe out of its
preferred direction, reducing the overlap between the orbitals
that participate in the hopping process, and therefore to the
observed enhancement of the resistance in a magnetic field.

Applying a magnetic field along the easy axis allows the Fe $e_g$
orbital to remain in its favored direction and so the overlap
between the $e_g$ orbital of Fe and the orbital of S is not
changed significantly. In this case the magnetoresistance remains
positive to lower temperatures, as it is shown in
figure~\ref{MReasy}. There the magnetoresistance of
Fe$_{0.95}$Cu$_{0.05}$Cr$_2$S$_4$ and
Fe$_{0.9}$Cu$_{0.1}$Cr$_2$S$_4$ is displayed with the magnetic
field applied along the easy and along the hard axis. When
applying the field along the easy axis, the magnetoresistance
exhibits a weak maximum. It changes sign at significantly lower
temperatures than upon application of the field along the hard
axis only for the sample with Cu concentration $x = 0.1$. This
change of sign may result from small misorientations of the sample
in the magnetic field, due to the experimental conditions.

\subsection{Influence of hydrostatic pressure}
\begin{figure}
\includegraphics[clip,width=80mm]{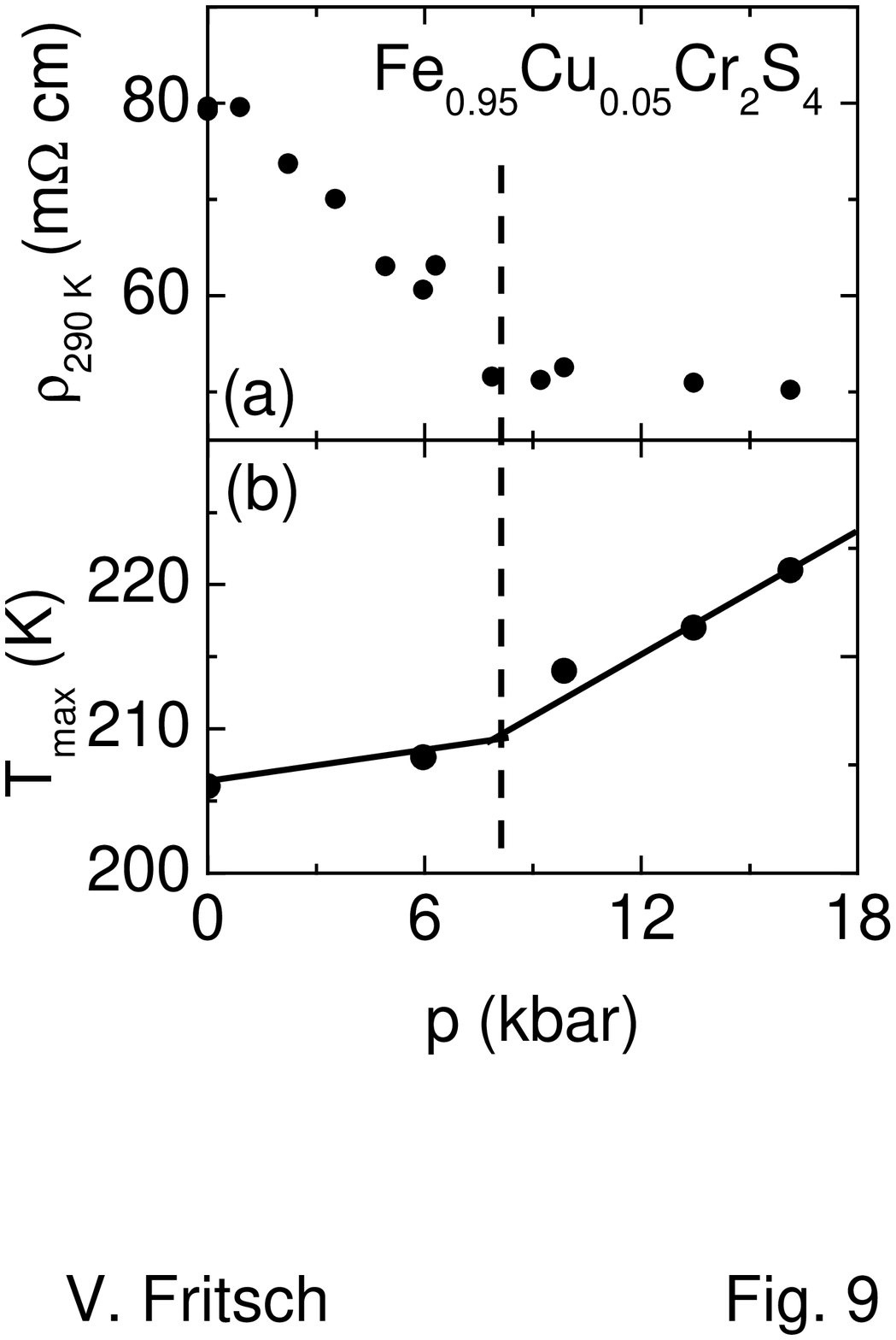}
\caption{(a) Resistivity of Fe$_{0.95}$Cu$_{0.05}$Cr$_2$S$_4$ at
$T = 290$~K in dependence of the applied hydrostatic pressure. (b)
Shift of the local maximum in the resistivity of
Fe$_{0.95}$Cu$_{0.05}$Cr$_2$S$_4$, as obtained from
figure~\ref{Fe50V}~(b)(closed circles, $\bullet$). The dashed line
indicates the critical pressure at about $p \approx 8$~kbar.
\label{vvsp}}
\end{figure}

In Ref.~\onlinecite{tsu01b} was shown that by the application of
pressure the Curie temperature $T_{\mathrm{C}}$ is shifted to
higher temperatures as indicated by the shift of the temperature
of the local maximum $T_{\mathrm{max}}$ of the $\rho (T)$ curves.
Therefore we conclude that the same effect works for the Cu-doped
compounds, and the shift of $T_{\mathrm{max}}$ can be taken as the
shift of $T_{\mathrm{C}}$.

The pressure dependence of the resistivity of
Fe$_{0.95}$Cu$_{0.05}$Cr$_2$S$_4$ at $T = 290$~K is drawn in
figure~\ref{vvsp}~(a). At a pressure of approximately $8$~kbar the
resistivity has declined about $37\%$ from its value at ambient
pressure. For higher pressures $\rho (290~\mathrm{K})$ stays
constant. On the other hand, the temperature of the local maximum
in the $\rho(T)$ curve (see figure~\ref{Fe50V}~(b)) increases
stronger with pressures above $8$~kbar, as shown in
figure~\ref{vvsp}~(b). Therefore one can assume that the effect of
hydrostatic pressure is changed, when a critical value $p \approx
8$~kbar is exceeded. In contrast to La$_{1-x}$Sr$_{x}$MnO$_3$,
where a linear pressure dependence was found,\cite{mor95} in the
present system two different pressure regimes with different
pressure gradients in $T_{\mathrm{max}}$ are in place.

\begin{figure}
\includegraphics[clip,width=80mm]{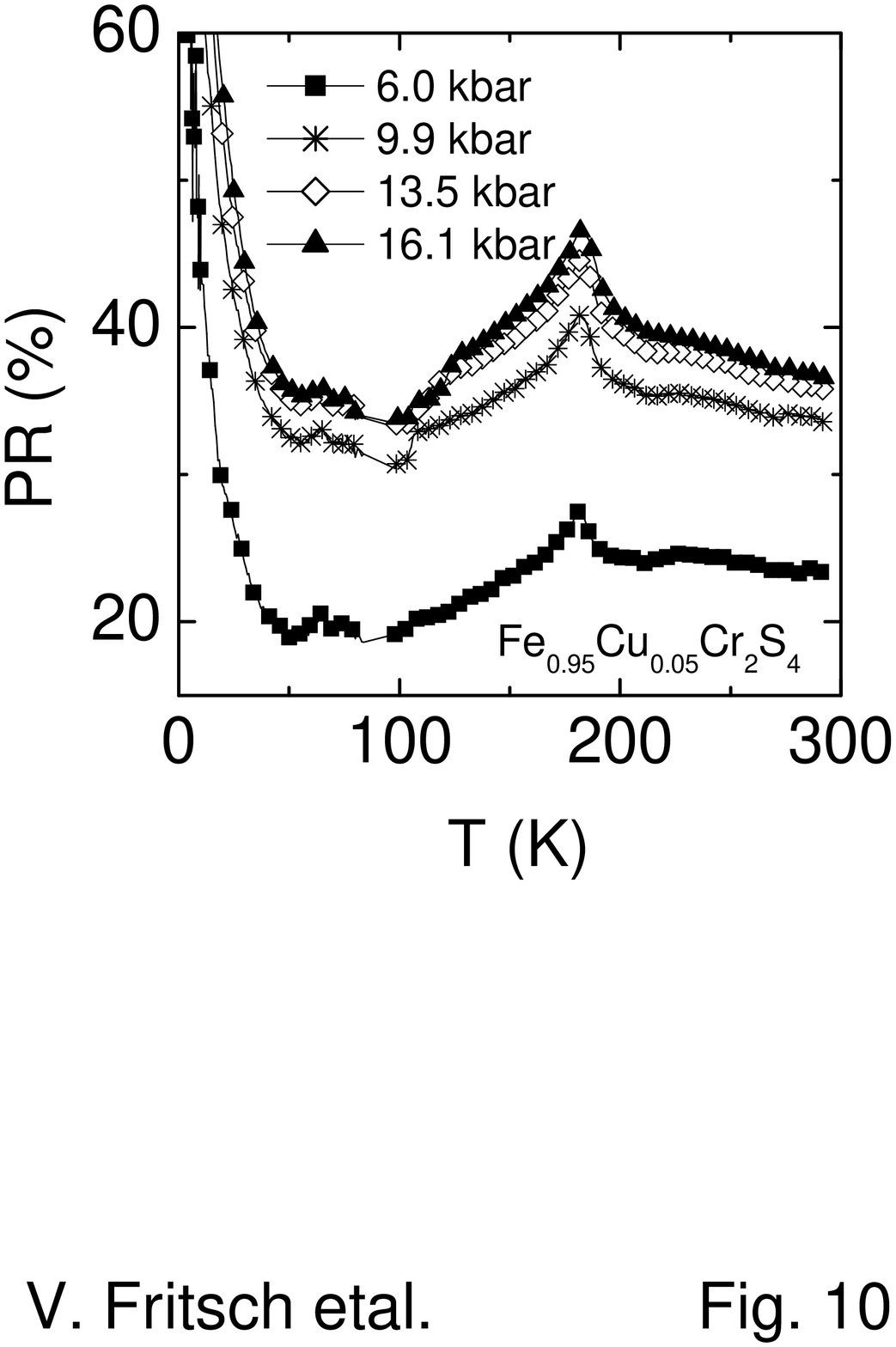}
\caption{Relative change of resistivity under hydrostatic pressure
$PR = \frac{\rho(0 \mathrm{kbar}) - \rho (p)}{\rho(0
\mathrm{kbar})}$. \label{PFe50V}}
\end{figure}

Figure~\ref{PFe50V} displays the effect of pressure on the
electrical resistance ($PR$), which is defined in analogy to the
magnetoresistance $MR$ as $PR := \frac{\rho(0~\mathrm{kbar}) -
\rho(p)}{\rho(0~\mathrm{kbar})}$. There are two remarkable
features: first of all, at the Curie temperature $T_{\mathrm{C}}$
a peak, similar to the magnetoresistance, arises, however, second
the value of $PR$ does not change sign at low temperatures and its
absolute value increases up to $100~\%$.

The application of hydrostatic pressure is expected to increase
the overlap between the orbitals and to broaden the bands,
resulting in an enhanced mobility of the charge carriers and a
reduction of the energy gap between the bands. This yields an
enhanced electric conductivity, which is illustrated in
figure~\ref{PFe50V}. Similar behavior was found in manganites, for
example in polycrystalline La$_{1-x}$Ca$_x$MnO$_3$.\cite{neu95} If
one would approximately describe the different conducting
mechanisms with different hopping laws, a reduction of the hopping
barriers automatically yields the strong increase of the $PR$
value at low temperatures. However, it is necessary to bear in
mind that the pressure is relatively moderate in the present
study. Thus its effect on the hopping barriers is not expected to
be such large and one has to look for an other mechanism. For
example the pressure might affect the Jahn-Teller distortion and
this way reinforce the conductivity.

\section{Conclusion}
In this paper X-ray, magnetization, FMR and resistivity data from
single crystals of Fe$_{1-x}$Cu$_{x}$Cr$_2$S$_4$ are presented.
The results are discussed in a hopping model, where the
conductivity is explained by triple-exchange mechanisms for the
concentration range $x < 0.5$ and double-exchange for $x \geq
0.5$.

Applying an external magnetic field or hydrostatic pressure to the
system ($x \leq 0.5$) has qualitatively an analogous effect for
temperatures around the Curie temperature $T_{\mathrm{C}}$: the
overlap of the orbitals is enhanced and the bands are broadened.
Thus the conductivity increases, while $T_{\mathrm{C}}$ is shifted
upward. At lower temperatures this similarity of the effect of an
external magnetic field and hydrostatic pressure vanishes. While
the value of $PR$ shows a strong upturn at low temperatures, in
the magnetoresistance a strong anisotropy arises. Applying the
magnetic field along the hard axis leads to a strong negative
magnetoresistance, while applying the field along the easy axis
results in a flat maximum in the magnetoresistance. Since the
origin of this unusual feature is still unclear, further
investigations of the electronic and orbital correlations in
Fe$_{1-x}$Cu$_x$Cr$_2$S$_4$ are needed and a promising challenge
for future experiments and theoretical calculations.

\acknowledgments We would like to thank V. Sidorov for his
assistance in the pressure measurements. This work was supported
by the BMBF via VDI/EKM, FKZ 13N6917/18 and by DFG within SFB 484
(Augsburg). Work at Los Alamos was performed under the auspices of
the U.S. DOE.

\end{document}